\documentclass[aps,prl,twocolumn,groupedaddress]{revtex4}
\usepackage[dvips]{graphics, color}

\begin{document}

\title{Creation of ultracold molecules from a Fermi gas of atoms}

\author{C. A. Regal, C. Ticknor, J. L. Bohn, and D. S.
Jin\cite{adr1}}

\affiliation{JILA, National Institute of Standards and Technology
and Department of Physics, University of Colorado, Boulder, CO
80309-0440}

\date{\today}

\begin{abstract} Since the realization of Bose-Einstein
condensates (BEC) in atomic gases an experimental challenge has
been the production of molecular gases in the quantum regime.  A
promising approach is to create the molecular gas directly from an
ultracold atomic gas; for example, atoms in a BEC have been
coupled to electronic ground-state molecules through
photoassociation \cite{Wynar} as well as through a magnetic-field
Feshbach resonance \cite{Donley}. The availability of atomic Fermi
gases provides the exciting prospect of coupling fermionic atoms
to bosonic molecules, and thus altering the quantum statistics of
the system. This Fermi-Bose coupling is closely related to the
pairing mechanism for a novel fermionic superfluid proposed to
occur near a Feshbach resonance \cite{Holland,Timmermans}.  Here
we report the creation and quantitative characterization of
exotic, ultracold $^{40}$K$_2$ molecules. Starting with a quantum
degenerate Fermi gas of atoms at $T < 150$ nanoKelvin we scan over
a Feshbach resonance to adiabatically create over a quarter
million trapped molecules, which we can convert back to atoms by
reversing the scan. The small binding energy of the molecules is
controlled by detuning from the Feshbach resonance and can be
varied over a wide range. We directly detect these weakly bound
molecules through rf photodissociation spectra that probe the
molecular wavefunction and yield binding energies that are
consistent with theory. \end{abstract}

\maketitle

Scattering resonances known as Feshbach resonances occur when the
collision energy of two free atoms coincides with that of a
quasi-bound molecular state \cite{Feshbach,Stwalley,Tiesinga}. By
varying the strength of an external magnetic field experimenters
can tune the relative atom-molecule energy through the Zeeman
effect. This enables control over the strength of cold atom
interactions, characterized by the s-wave scattering length $a$,
as well as whether they are effectively repulsive $(a>0)$ or
attractive $(a<0)$. These resonances have been used to tune the
interaction strength between atoms for both Bose and Fermi gases
\cite{Inouye,Cornish,Loftus,Ketterle,Thomas,Regal,Ohara,meanfield,Salomon}.

Another possible use of a Feshbach resonance is to controllably
convert atoms into molecules. The energy of the molecular state
associated with the Feshbach resonance coincides with the free
atom threshold at the resonance peak, and the binding energy of
the molecule varies smoothly with magnetic field on the repulsive
side of the resonance $(a>0)$.  Thus, one would expect that atoms
could be coupled to molecules by ramping the detuning from the
resonance using a time-dependent magnetic field
\cite{Timmermans2,Abeelen,Mies}. For example, ramping the magnetic
field through a Feshbach resonance resulted in large losses in a
$^{23}$Na BEC \cite{Stenger}. Further, coherent oscillations
between atoms and molecules in a BEC were observed following
magnetic field pulses on the repulsive side of a $^{85}$Rb
Feshbach resonance \cite{Donley}.  Here we report the efficient
creation of bosonic molecules from fermionic $^{40}$K atoms using
a magnetic field ramp across a Feshbach resonance. We present
clear evidence of diatomic molecule formation through direct,
spectroscopic detection of these molecules.

\begin{figure} \begin{center}
\scalebox{.6}[.6]{\includegraphics{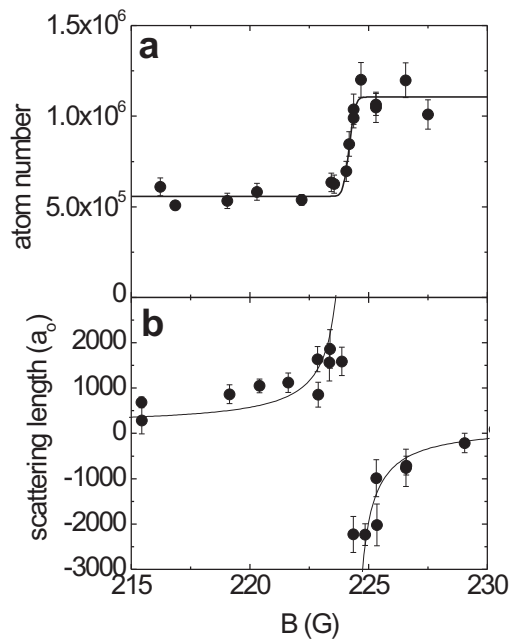}} \caption{(a) Loss
of $|9/2, -9/2 \rangle$  and $|9/2, -5/2 \rangle$ atoms as a
function of the final magnetic field value $B$ of a $40$ $\mu {\rm
s/G}$ ramp that starts at 227.81 G. The atom cloud is initially at
$T/T_F=0.21$ and $n_{pk}=2.1 \times 10^{13}$, where $n_{pk}$ is
the peak atomic density of the two-component cloud. The data are
fit to an error function. The resulting gaussian width of the
transition region is $\sigma=0.21 \pm 0.07$ G, and the transition
position is $B_0=224.18 \pm 0.05$ G. (b) For comparison we plot a
previous measurement of the Feshbach resonance, where the
resonance peak was found to be at $224.21 \pm 0.05$ G
\cite{meanfield}.} \label{Feshbach} \end{center} \end{figure}

The experiments reported here employ previously developed
techniques for cooling and spin state manipulation of $^{40}$K
\cite{Loftus,DeMarco,meanfield}. Because of the quantum statistics
of fermions a mixture of two components, for example atoms in
different internal spin states, is required to have s-wave
interactions in the ultracold gas. With a total atomic spin
$f=9/2$ in its lowest hyperfine ground state, $^{40}$K has ten
available Zeeman spin-states $|f,m_f\rangle$, where $m_f$ is the
spin projection quantum number. Mixtures of atoms in two of these
spin states are used in evaporative cooling of the gas, first in a
magnetic trap and then in a far-off resonance optical dipole trap.
The optical trapping potential has the distinct ability to trap
atoms in any spin state as well as any molecules created from
these atoms. For these experiments the optical trap is
characterized by radial frequencies ranging from $\nu_r = 215$ to
$276$ Hz, with the trap aspect ratio, $\nu_r/\nu_z$, fixed at $70
\pm 20 $. The temperature $T$ of these two-component gases,
measured in units of the Fermi temperature $T_F$, ranges from
$T/T_F=0.13$ to $0.33$. This degree of quantum degeneracy is near
the lowest ever demonstrated in a Fermi gas of atoms
\cite{meanfield}.

Experiments are initiated by preparing atoms in a nearly equal,
incoherent mixture of the $|9/2,-5/2\rangle$ and
$|9/2,-9/2\rangle$ spin states. With these states we access a
previously reported Feshbach resonance located at a magnetic field
of $224.21 \pm 0.05$ G (Fig.~\ref{Feshbach}b) \cite{meanfield}. A
time-dependent current through an auxiliary coil provides magnetic
field ramps near the resonance. Starting from a magnetic field of
$227.81$ G, the field is ramped at a rate of $(40$ $\mu {\rm
s/G})^{-1}$ across the resonance to various final values.  The
number of atoms remaining following the ramp is determined from an
absorption image of the cloud (at $\sim 4$ G) after expansion from
the optical trap. Since the light used for these images is
resonant with the atomic transition, but not with any molecular
transitions, we selectively detect only the atoms.   In
Fig.~\ref{Feshbach}a we present the observed atom number as a
function of the final magnetic field value of the ramp.  We find
that the atoms disappear abruptly at the Feshbach resonance peak.

\begin{figure} \begin{center}
\scalebox{.8}[.8]{\includegraphics{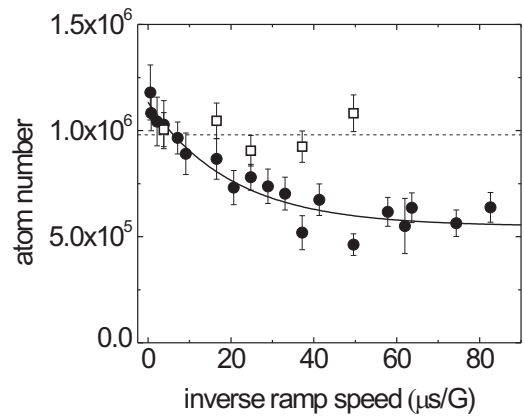}}\caption{Atom
number following magnetic field ramps across the Feshbach
resonance from high to low field (circles).  The data are taken
with $T/T_F = 0.33$ and $n_{pk}=1.4 \times 10^{13}$ ${\rm
cm}^{-3}$. In accordance with a Landau-Zener model, we fit the
data to an exponential (solid line); we find that the decay
constant is $20 \pm 6$ $\mu$s/G. Data taken at $T/T_F = 0.13$ and
$n_{pk}=9 \times 10^{12}$ ${\rm cm}^{-3}$ behave similarly.  The
squares represent data for which the magnetic field was ramped
first at a rate of $(40$ $\mu {\rm s/G})^{-1}$ across the
resonance and then ramped back above the Feshbach resonance at
varying rates.} \label{ramp} \end{center} \end{figure}

We also investigate the atom loss as a function of the rate of the
magnetic field sweep.  Fig.~\ref{ramp}~illustrates the result of
linear magnetic field ramps across the Feshbach resonance from
$228.25$ G to $216.15$ G. For our fastest sweeps there is no
observable effect upon the atoms, while significant atom number
loss is observed for slower sweeps.  For our slowest sweeps we
find that the number of atoms lost saturates at $50\%$. The atoms
vanish at least two orders of magnitude more quickly than expected
from previously measured inelastic collision rates at a resonance
\cite{Regal}. When we reverse the process by applying an
additional magnetic field ramp across the Feshbach resonance in
the opposite direction, we observe a return of nearly all the
``lost'' atoms (Fig.~\ref{ramp}). This is consistent with the lost
atom number corresponding to trapped molecules.

Surprisingly, the number of molecules produced is very large
despite the fact that the Fermi gas is not described by a single
wavefunction as is a BEC. In fact the number of molecules is
sufficiently large that the temperature of our initial atomic gas
is below the molecular Bose-Einstein condensation temperature in
the optical trap. Further, we can measure the lifetime $\tau$ of
the molecules by varying the time before conversion back to atoms;
we find that $\tau \sim 1$ ms.

\begin{figure} \begin{center}
\scalebox{0.5}[0.5]{\includegraphics{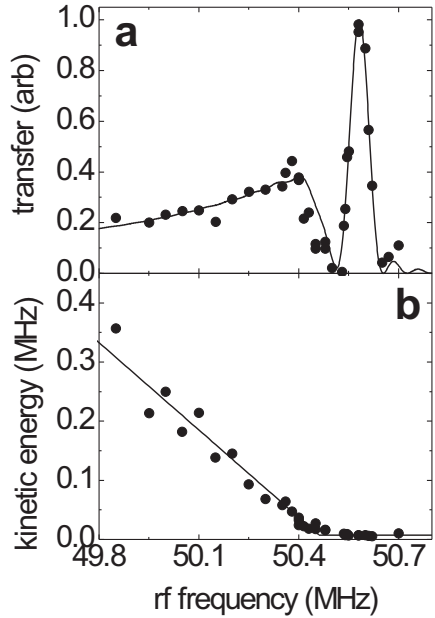}} \caption{(a)
Dissociation spectrum at $B_{hold}=222.49$ G, with an original
atom cloud containing $N=1.4 \times 10^6$ atoms at $T/T_F = 0.26$,
where $N$ is the total number of atoms in the two-component cloud.
The number of $|9/2,-7/2\rangle$ atoms after an applied rf pulse
is plotted versus rf frequency. The solid line is the weighted sum
of the atomic transition line and the calculated dissociation
spectrum, both convolved with the frequency width of the rf pulse.
(b) Resulting $|9/2,-7/2\rangle$ atom kinetic energy. Two separate
linear fits are applied to the data to determine the threshold
position. The slope beyond threshold is $0.49 \pm 0.03$; this
indicates that the atom pair ($|9/2,-7/2\rangle$ +
$|9/2,-9/2\rangle$) receives the additional energy $\Delta E$
beyond the binding energy when the molecule is dissociated.}
\label{spectrum} \end{center} \end{figure}

Using radio frequency spectroscopy we directly probe these
molecules. First, we create the molecules with a $(40$ $\mu {\rm
s/G})^{-1}$ magnetic field ramp. This ramp starts of $227.81$ G
and ends at various final magnetic field values $B_{hold}$ below
the resonance. At $B_{hold}$ a 13 $\mu$s rf pulse is applied to
the cloud; the rf frequency is chosen so that the photon energy is
near the energy splitting between the $|9/2,-5/2\rangle$ and $
|9/2,-7/2\rangle$ atom states.  The resulting population in the
$|9/2,-7/2\rangle$ state, which is initially unoccupied, is then
probed selectively either by separating the spin states spatially
using a strong magnetic field gradient during free expansion
(Stern-Gerlach imaging), or by leaving the magnetic field high
(215 G) and taking advantage of nonlinear Zeeman shifts.

\begin{figure}[ht] \begin{center}
\scalebox{1}[1]{\includegraphics{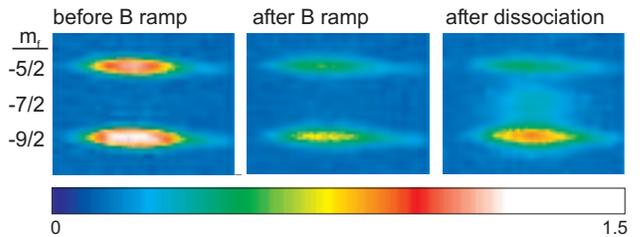}}
\caption{Absorption images of the quantum gas using a
Stern-Gerlach technique. We start with ultracold fermionic atoms
in the $|9/2,-5/2\rangle$ and $|9/2,-9/2\rangle$ states of
$^{40}$K. A magnetic field ramp through the Feshbach resonance
causes 50$\%$ atom loss, due to adiabatic conversion of atoms to
diatomic molecules.  To directly detect these bosonic molecules we
apply an rf photodissociation pulse; the dissociated molecules
then appear in the $|9/2,-7/2\rangle$ and $|9/2,-9/2\rangle$ atom
states. The shaded bar indicates the optical depth.}
\label{pictures} \end{center} \end{figure}

Fig.~\ref{spectrum}a shows a sample rf spectrum at
$B_{hold}=222.49$ G; the resulting number of atoms in the
$|9/2,-7/2\rangle$ state is plotted as a function of the frequency
of the rf pulse.  We observe two distinct features in the
spectrum.  The sharp, symmetric peak is very near the expected
$|9/2,-5/2\rangle$ to $|9/2,-7/2\rangle$ transition frequency for
free atoms $\nu_{atom}$. With the Stern-Gerlach imaging we see
that the total number of $|9/2,-5/2\rangle$ and $|9/2,-7/2\rangle$
atoms is constant, consistent with transfer between these two atom
states. The width of this line is defined by the fourier width of
the applied rf pulse. Nearby is a broader, asymmetric peak shifted
lower in frequency. Here we find that after the rf pulse the total
number of observed atoms ($|9/2,-5/2\rangle$ $+$
$|9/2,-7/2\rangle$) actually increases (Fig.~\ref{pictures}).
Also, the resulting $|9/2,-7/2\rangle$ gas in this region has a
significantly increased kinetic energy, which grows linearly for
larger frequency shifts from the atom peak (Fig.~\ref{spectrum}b).

The asymmetric peak can be interpreted as the dissociation of
molecules into free $|9/2,-7/2\rangle$ and $|9/2,-9/2\rangle$
atoms.  Since the applied rf pulse stimulates a transition to a
lower energy Zeeman state, we expect $h \nu_{rf}=h \nu_{atom}
-E_{binding}-\Delta E$, where $h$ is Planck's constant,
$E_{binding}$ is the binding energy of the molecule, and we have
ignored mean-field interaction energy shifts. The remaining
energy, $\Delta E$, must by imparted to the dissociating atom pair
as kinetic energy. Further, the observed lineshape of the
asymmetric peak should depend on the Franck-Condon factor, which
gives the overlap of the molecular wavefunction with the atomic
wavefunction.

We have calculated this multichannel Franck-Condon overlap as a
function of energy. The resulting transition rate, convolved with
the frequency width of the applied rf and scaled vertically, is
shown as the solid line in Fig.~\ref{spectrum}a. The agreement
between theory and experiment for the dissociation spectrum is
quite good. This well-resolved spectrum provides much information
about the molecular wavefunction.  For example, the mean
interatomic separation of the molecules at this magnetic field is
extremely large, $\sim 170$ $a_0$, where $a_0$ is the Bohr radius.

\begin{figure}[h] \begin{center}
\scalebox{.95}[.95]{\includegraphics{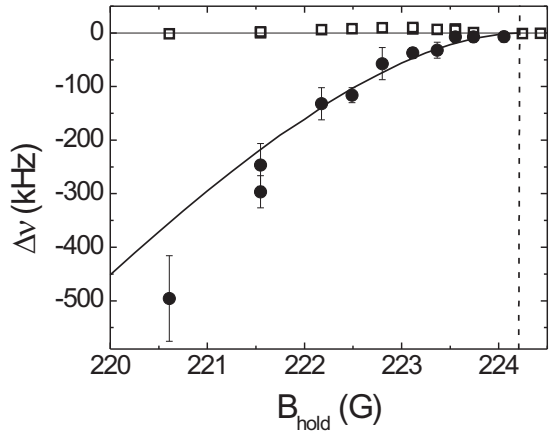}}
\caption{Binding energy of the molecules.  The frequency shift
$(\Delta \nu)$ from the expected $|9/2,-5/2\rangle \rightarrow
|9/2,-7/2\rangle$ transition is plotted versus magnetic field for
the $|9/2,-7/2\rangle$ atoms (squares) and the molecules
(circles). The typical atom cloud before molecule creation is
characterized by $T/T_F = 0.14$ and $N=7 \times 10^5$. The dashed
line indicates the Feshbach resonance position. The solid line is
a calculation of the binding energy of the molecules as a function
of detuning from the resonance. The small shift in the atom
transition frequency is due to the atom-atom interaction energy.}
\label{frequency} \end{center} \end{figure}

In Fig.~\ref{frequency} we plot the magnetic field dependence of
the frequency shift $\Delta \nu$ between the atom line and the
threshold of the molecular spectrum, which should correspond to
the molecular binding energy. While this could in principle be
obtained directly from the transfer spectrum
(Fig.~\ref{spectrum}a), the appearance of the threshold in the
energy of the $|9/2,-7/2\rangle$ cloud is more clear
(Fig.~\ref{spectrum}b). We compare the position of this energy
threshold to the expected atom-atom transition frequency
$\nu_{atom}$ based upon a calibration of the magnetic field
strength. The data are consistent with a theoretical calculation
of the binding energy (solid line) based upon a full coupled
channels calculation with no free parameters.

This agreement with theory leaves no doubt that we are creating
large numbers of weakly bound molecules.  These highly
vibrationally excited molecules could be used to study ultracold
molecule-molecule or molecule-atom collisions
\cite{Balakrishmana,Forrey,Soldan}.  Further, the explicit
coupling of a quantum degenerate gas of Fermi atoms to bosonic
molecules could possibly be developed as a method to facilitate
creation of a predicted exotic fermionic superfluid. In addition
our molecule detection technique could be extended to measure the
gap energy in this superfluid phase \cite{Petrosyan,Torma}.

Acknowledgements:  We thank E. A. Cornell, C. E. Wieman, C. H.
Greene, and S. Inouye for useful discussion. This work was
supported by NSF and NIST, and C. A. R. acknowledges support from
the Hertz Foundation.

\end{document}